%% file: v9.tex
\documentclass[prb, preprint]{revtex4-1}
\usepackage{framed}
\usepackage{graphicx}
\usepackage{dcolumn}
\usepackage{amsmath}
\usepackage{amssymb}
\usepackage{subfigure,amsmath,verbatim,moreverb,bm}
\usepackage{color}
\def\up{\uparrow}
\def\down{\downarrow}
\def\be{\begin{equation}}
\def\ee{\end{equation}}
\def\ber{\begin{eqnarray}}
\def\eer{\end{eqnarray}}

\def\rv{{\bf r}}

\def\kv{{\bf k}}

\def\nn{\nonumber}

\def\Acalv{{\boldsymbol {\mathcal A}}}
\def\Acal{{\cal A}}

\begin{document}
\title{Spin Hall and Edelstein effects in metallic films: from 2D to 3D}
\author{J.Borge$^1$}
\author{C. Gorini$^2$}
\author{G. Vignale$^3$}
\author{R. Raimondi$^1$}
\affiliation{$^1$Dipartimento di Matematica e Fisica, 
Universit\`a  Roma Tre, Via della Vasca Navale 84, Rome, Italy}
\affiliation{$^2$Service de Physique de l'\'{E}tat Condens\'{e},
                     CNRS URA 2464, CEA Saclay,
                     F-91191 Gif-sur-Yvette, France}
\affiliation{$^3$Department of Physics and Astronomy, University of Missouri,  Columbia MO 65211, USA}

\begin{abstract}
A normal metallic film sandwiched between two insulators may have strong spin-orbit coupling 
near the metal-insulator interfaces, even if spin-orbit coupling is negligible in the bulk of the film.
In this paper we study two technologically important and deeply interconnected effects that arise 
from interfacial spin-orbit coupling in metallic films.  The first is the spin Hall effect, 
whereby a charge current in the plane of the film is partially converted into an orthogonal spin current 
in the same plane.  The second is the Edelstein effect, in which a charge current produces an
in-plane, transverse spin polarization.  At variance with strictly two-dimensional Rashba systems, 
we find that the spin Hall conductivity has a finite value even if spin-orbit interaction with impurities 
is neglected and ``vertex corrections'' are properly taken into account.
Even more remarkably, such finite value becomes ``universal'' in a certain configuration.  
This is a direct consequence of the spatial dependence of spin-orbit coupling on the third dimension, 
perpendicular to the film plane.  The non-vanishing spin Hall conductivity has a profound influence 
on the Edelstein effect, which we show to consist of two terms, the first with the standard
form valid in a strictly two-dimensional Rashba system, and a second arising from the presence
of the third dimension. 
Whereas the standard term is proportional to the momentum relaxation time, 
the new one scales with the spin relaxation time.  
Our results, although derived in a specific model, should be valid rather generally, whenever
a spatially dependent Rashba spin-orbit coupling is present and the electron motion 
is not strictly two-dimensional.

 \end{abstract}
\maketitle

\section{Introduction}
\label{sec_intro}

Spin-orbit coupling gives rise to several interesting transport phenomena arising from the 
induced correlation between charge and spin degrees of freedom.  
In particular, it allows one to manipulate spins without using magnetic electrodes, 
having as such become one of the most studied topics within the field of spintronics. 
\cite{Hirsch99, Zhang00, Murakami03, Sinova04, Handbook,
Raimondi_2Dgas_PRB06, Culcer_SteadyState_PRB07, Culcer_Generation_PRL07, Culcer_SideJump_PRB10, TsePRB05,
GalitskiPRB06, Tanaka_NJP09,Hankiewicz09,TenYears2010}
Among the many interesting effects that arise from spin-orbit coupling, two stand out 
for their potential technological importance: the spin Hall effect\cite{Dyakonov71} 
and the  Edelstein effect\cite{Lyanda-Geller89,Edelstein1990}.
The spin Hall effect consists in the appearance of a $z$-polarized spin current flowing in the $y$-direction
produced by an electric field in the $x$-direction.\cite{Kato04, Sih05, Wunderlich05, Stern06, Stern08}
The generation of a perpendicular electric field by an injected spin current,
i.e. the inverse spin Hall effect, has been observed in numerous settings and presently provides the basis 
for one of the most effective methods to detect spin currents.\cite{Valenzuela_Nat06,Takahashi_Revese_PRL07,Takahashi_GSH_NatMater08}
The Edelstein effect\cite{Lyanda-Geller89,Edelstein1990} consists instead in the appearance of a $y$-spin
polarization in response to an applied electric field in the $x$-direction.
It has been proposed as a promising way
of achieving all-electrical control of magnetic properties in electronic
circuits.\cite{Ioan2010,Kato04,Sih05,Inoue2003,Yang2006,Chang2007,Koehl2009,Kuhlen2012}
The two effects are deeply connected,\cite{Gorini08,Raimondi09,Shen2013} as we will see momentarily.

There are, in principle, several possible mechanisms for the spin Hall effect,
and it is useful to divide them in two classes.  We call them either extrinsic or intrinsic,
depending on whether their origin is the spin-orbit interaction with impurities
or with the regular lattice structure.  In this work we will focus exclusively on intrinsic effects.  
This means that the impurities (while, of course, needed to give the system a finite electrical conductivity) 
do not couple to the electron spin.

Bychkov and Rashba devised an extremely simple and yet powerful model\cite{Rashba84}
describing the intrinsic spin-orbit coupling of the electrons in a 2-Dimensional Electron Gas (2DEG) 
in a quantum well in the presence of an electric field perpendicular to the plane in which the electrons move.
In spite of its apparent simplicity, this analytically solvable model has several subtle features, 
which arise from the interplay of spin-orbit coupling and impurity scattering.  
The best-known feature is the vanishing of the Spin Hall Conductivity (SHC) for a uniform and constant 
in-plane electric field.\cite{Mishchenko2004,Raimondi05,Khaetskii2006}
This would leave spin-orbit coupling with impurities (not included in the original Bychkov-Rashba model) 
as the only plausible mechanism for the experimentally observed spin Hall effect 
in semiconductor-based 2DEGs~\cite{Kato04,Sih05}. 
\footnote{An alternative source of intrinsic spin Hall effect is random fluctuations of the electric field 
perpendicular to the 2DEG - see Dugaev et al.~\cite{Dugaev_PRB10}.}

\begin{figure}
\begin{center}
\includegraphics[width=3in]{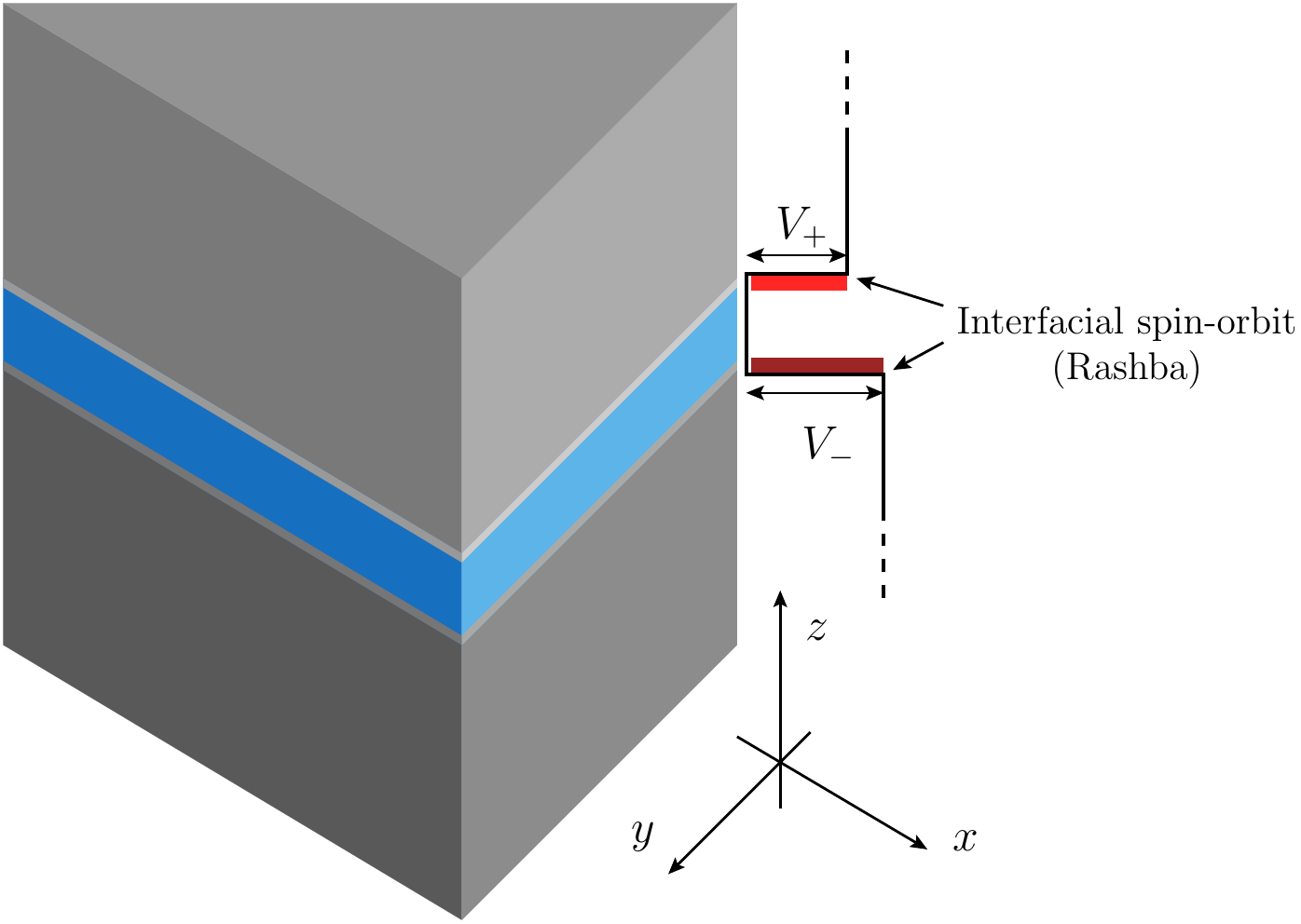}
\caption{(color online) Schematic representation of a thin metal film sandwiched between insulators 
with asymmetric interfacial spin-orbit couplings.  $V_+$ and $V_-$ are the heights of the two 
interfacial potential barriers.  These potentials generate interfacial spin-orbit interactions 
of the Rashba type, whose strength is controlled by the effective Compton wavelengths 
$\lambda_+$ and $\lambda_-$ respectively.}
\label{setup}
\end{center}
\end{figure}

However it has been recently pointed out that the vanishing of the SHC need not occur
in systems which are not strictly two-dimensional, as explicitly shown
in a model schematically describing the interface of the two insulating oxides LaAlO$_3$ and
SrTiO$_3$ (LAO/STO)\cite{Hayden13}. 
Even more recently~\cite{Wang2013}, it has been suggested that a large SHC could be realized 
in a thin metal (Cu) film that is sandwiched between two different insulators, 
such as oxides or the vacuum.\footnote{A very large spin Hall angle of extrinsic origin has been 
observed~\cite{Niimi2012} in thin films of Cu doped with bismuth impurities. 
In Ref.~\onlinecite{Wang2013}, however, the Bi impurities are absent.}  
Such a system is shown schematically in Fig.~\ref{setup}.
The inversion symmetry breaking across the interfaces produces interfacial Rashba-like spin-orbit couplings,
thus allowing metals without substantial intrinsic bulk spin-orbit to host a non-vanishing SHC.
The spin-orbit coupling asymmetry -- or, more generally, the fact that the spin-orbit interaction
is not homogeneous across the thickness of the film -- is the core issue in this novel approach.
In this paper we will study the influence of the interfacial spin-orbit couplings on the Edelstein 
and spin Hall effects in this class of heterostructures.

Before proceeding to a detailed study of the model depicted in Fig.~\ref{setup},
it is useful to recall the deep connection\cite{Gorini08,Raimondi09,Shen2013}
that exists between the spin Hall and Edelstein effects in the Bychkov-Rashba model, described by the Hamiltonian
\be
\label{RASHBA}
H=\frac{p^2}{2m}+\alpha (\sigma_xp_y-\sigma_yp_x)\,,
\ee
where $m$ is the effective electron mass and $\alpha$ is the Bychkov-Rashba spin-orbit coupling constant
given by $\alpha=\lambda^2 eE_z/\hbar$, with $\lambda$ the materials' effective Compton wavelength,
$E_z$ the electric field perpendicular to the electron layer, and $e$ the absolute value of the electron charge.
It is convenient to describe spin-orbit coupling in terms of a non-Abelian gauge field
$\Acalv=\Acal^a\sigma^a/2$, with $\Acal^x_y= 2m\alpha$ and $\Acal^y_x=-2m\alpha$.
\cite{Tokatly_Color_PRL08,Gorini10,Takeuchi2013}
If not otherwise specified, superscripts indicate spin components, while subscripts stand for spatial components.
The first consequence of resorting to this language is the appearance of an $SU(2)$ magnetic field
${\cal B}^z_z=-(2m\alpha)^2$, which arises from the non-commuting components of the Bychkov-Rashba vector potential.
Such a spin-magnetic field couples the charge current driven by an electric field, say along $x$,
to the $z$-polarized spin current flowing along $y$.  This is very much similar to the standard Hall effect,
where two {\it charge} currents flowing perpendicular to each other are coupled by a magnetic field.
The drift component of the spin current can thus be described by a Hall-like term
\be
\label{SHDRIFT}
[J^z_y]_{drift}=\sigma^{SHE}_{drift} E_x.
\ee
It is however important to appreciate that this is not yet the full spin Hall current, 
i.e. $\sigma^{SHE}_{drift}$ is not the full SHC.  In the diffusive regime $\sigma^{SHE}_{drift}$ is given 
by the classic formula $\sigma^{SHE}_{drift}=(\omega_c \tau)\sigma_D/e$, where $\omega_c = {\cal B}/m\hbar $ 
is the ``cyclotron frequency'' associated with the $SU(2)$ magnetic field, 
$\tau$ is the elastic momentum scattering time, and $\sigma_D$ is the Drude conductivity.  
For a more general formula see Eq.~(\ref{discussion_5}) below.

In addition to the drift current, there is also a ``diffusion current'' due to spin precession 
around the Bychkov-Rashba effective spin-orbit field.
Within the $SU(2)$ formalism this current arises from the replacement of the ordinary derivative with the 
$SU(2)$ covariant derivative in the expression for the diffusion current.  The $SU(2)$ covariant derivative, 
due to the gauge field, is
\be
\nabla_j {\cal O} = \partial_j {\cal O} + i\left[\Acal_j,{\cal O}\right],
\ee
with ${\cal O}$ a given quantity being acted upon.  The normal derivative, $\partial_j$, along a given axis $j$
is shifted by the commutator with the gauge field component along that same axis.
As a result of the replacement $\partial \to \nabla$ diffusion-like terms,
normally proportional to spin density gradients,
arise even in uniform conditions and the diffusion contribution to the spin current turns out to be
\be
\label{DIFFUSION}
[J^z_y]_{diff}= \frac{2m\alpha}{\hbar} D s^y,
\ee
where $D=v_F^2\tau /2$ is the diffusion coefficient, $v_F$ being the Fermi velocity.  
In the diffusive regime the full spin current $J^z_y$ can thus be expressed as
\begin{equation}
\label{discussion_1}
J^z_y=\frac{2m\alpha}{\hbar} D s^y +\sigma_{drift}^{SHE}E_x.
\end{equation}
For a detailed justification of Eq.~\eqref{discussion_1} we refer the reader to
Refs.~\onlinecite{Gorini10,Raimondi_AnnPhys12}.
The factor in front of the spin density in the first term of Eq.(\ref{discussion_1})
can also be written as an effective velocity $L_{so}/\tau_s$.  Here $L_{so}=\hbar (2m \alpha )^{-1}$
is the typical spin length due to the different Fermi momenta in the two spin-orbit split bands,
whereas $\tau_s=\hbar^2(4m^2\alpha^2 D)^{-1}$ is the Dyakonov-Perel spin relaxation time.
In terms of $\tau$ and $\tau_s$ one has 
\begin{equation}
\label{discussion_5}
\sigma_{drift}^{SHE}=\frac{e}{8\pi\hbar }\frac{2\tau}{\tau_s}\,,
\end{equation}
which is indeed equivalent to the classical surmise given after Eq.~(\ref{SHDRIFT}).
If we introduce the total SHC and the Edelstein Conductivity (EC) defined by
\begin{equation}
\label{discussion_2}
J^z_y= \sigma^{SHE} E_x, \   s^y=\sigma^{EE}E_x
\end{equation}
we may rewrite Eq.(\ref{discussion_1}) as
\begin{equation}
\label{discussion_3}
\sigma^{EE}=\frac{\tau_s}{L_{so}} \left( \sigma^{SHE}-\sigma_{drift}^{SHE}\right).
\end{equation}
In the standard Bychkov-Rashba model a general constraint from the equation of motion dictates that
under steady and uniform conditions $J^z_y=0$.  Therefore the EC reads
\begin{equation}
\label{discussion_4}
\sigma^{EE}=-\frac{\tau_s}{L_{so}}\sigma_{drift}^{SHE}=-e \frac{m}{2\pi\hbar^2} \alpha \tau=-eN_0\alpha \tau,
\end{equation}
which is easily obtained by using the expressions given above and the single particle density of states
in two dimensions, $N_0=m/2\pi\hbar^2$.  The remarkable thing is that this expression remains unchanged
for arbitrary ratios between the spin splitting energy and the disorder broadening of the levels.
However, in a more general situation with a non-zero SHC the EC would consist of the two terms
appearing in Eq.~\eqref{discussion_3}.
The latter equation is the ``deep connection'' mentioned earlier between the Edelstein and the spin Hall effect.
The first term on the r.h.s. is the ``regular" contribution to the EC, 
the only surviving one in the Bychkov-Rashba model where the full SHC vanishes.  
The second term is  ``anomalous'' in the sense that it does not appear in the standard Bychkov-Rashba model, 
but it does appear in more general models such as the one we discuss in this paper.
Notice that the ``regular" term is proportional to $\tau$ (see Eq.~(\ref{discussion_5})),
while the ``anomalous"  term, being proportional to the Dyakonov-Perel relaxation time $\tau_s$  and, in the diffusive regime,
is {\it inversely} proportional to the momentum relaxation time.

At variance with the Bychkov-Rashba model, the one we choose for our system is not strictly two-dimensional, and
we take into account several states of quantized motion in the direction perpendicular to the interface ($z$).   
Another crucial feature of this model is the occurrence of two different spin-orbit couplings at the two interfaces.
The difference arises because (i) the interfacial potential barriers $V_+$ and $V_-$ are generally different, 
and (ii) the effective Compton wavelengths $\lambda_+$ and $\lambda_-$, characterizing the spin-orbit coupling
strength at the two interfaces, are different.

Our central results for the generic asymmetric model are
\begin{equation}
\label{SH4-Preview}
\sigma^{SHE}=-\sum_{n=1}^{n_c}\frac{e}{4\pi\hbar} \frac{\Delta E^{(3)}_{nk_{Fn}}}
{\Delta E_{nk_{Fn}}}\,,
\end{equation}
and
\be
\label{E9-Preview}
\sigma^{EE}=\sum_{n=1}^{n_c}\frac{eN_0}{k_{Fn}\hbar}\left[
\Delta E_{nk_{Fn}}\tau + \Delta E^{(3)}_{nk_{Fn}}\tau_{DP}^{(n)}
\right]\,,
\ee
the sums running over the $n_c$ filled $z$-subbands of the thin film.
To each subband there correspond a Fermi wavevector (without spin-orbit) $k_{Fn}$,
an intraband spin-orbit energy splitting with a linear- and a cubic-in-$k$ part
\ber
\label{deltaE}
\Delta E_{nk_{Fn}}
&=&
(2 E_0 n^2/d)[ k_{Fn}(\lambda_+^2-\lambda_-^2) +\frac{2 mk_{Fn}^3}{\hbar^2}(\lambda_+^6V_+-\lambda_-^6V_-)]
\\
\label{intra_split}
&\equiv&
\Delta E^{(1)}_{nk_{Fn}} + \Delta E^{(3)}_{nk_{Fn}}
\eer
and a Dyakonov-Perel spin relaxation time
\be
\label{a}
\frac{\tau_{DP}^{(n)}}{\tau}=2\left[
\frac{1+(2\tau \Delta E_{nk_{Fn}}/\hbar )^2}{(2\tau \Delta E_{nk_{Fn}}/\hbar)^2}
\right].
\ee
In the above formulas $d$ is the film thickness and $E_0=\hbar^2/2md^2$.
Two particularly interesting regimes are apparent.  First, a ``quasi-symmetric'' configuration,
defined by equal spin-orbit strengths, $\lambda_+=\lambda_-\equiv\lambda$, but different barrier heights, $V_+ \neq V_-$.
In this case $\Delta E^{(1)}_{nk}=0$ (due to Ehrenfest's theorem\footnote{This is because 
the splitting of the energy levels to first order in $k$ is shown by perturbation theory to be proportional 
to the expectation value of $V'(z)$, i.e. the force, in the ground-state in the absence of spin-orbit coupling.  
By Ehrenfest's theorem this is the expectation value of the time derivative of the $z$-component of the momentum, 
and therefore must vanish\cite{Winkler2003}.}) and a most striking result is obtained:
the SHC has a maximal value of $-\frac{e}{4\pi\hbar}$ (independent of $\lambda$!) 
times the number of occupied bands
\be
\sigma^{SHE} = -\sum_{n=1}^{n_c}\frac{e}{4\pi\hbar}.
\ee
At the same time the ``anomalous'' EC is at its largest.
A second very interesting configuration is a strongly asymmetric insulator-metal-vacuum
junction, $\lambda_+=0, V_+\rightarrow\infty$ and $\lambda_-\equiv\lambda, V_-\equiv V$.
In this case the SHC becomes directly proportional to the gap $V$
\be
\sigma^{SHE} = -\sum_{n=1}^{n_c}\frac{e}{4\pi\hbar^3}2mk^2_{Fn}V\lambda^4.
\ee
Notice however that the SHC cannot be made arbitrarily large simply by
engineering a large $V$, since the above result holds provided $2mk^2_{Fn}V\lambda^4/\hbar^2<1$.


The paper is organized as follows.  In Sec.~\ref{sec_model} we introduce and discuss the model.
In Secs.~\ref{sec_SHC} and ~\ref{sec_EC} we calculate the SHC and the EC, respectively.
Both Sections are technically heavy and can be skipped at a first reading, leading straight to
Sec.~\ref{sec_discussion} where the physical consequences of our results are discussed
and special regimes are analyzed.  Sec.~\ref{sec_conclusions} presents our summary and conclusions.




\section{The model and its solution}
\label{sec_model}

Following  Ref. \onlinecite{Wang2013}, we model the normal metallic thin film via the following Hamiltonian
\be\label{H1}
H=\frac{p^2}{2m}  + V_C(z) +H_R+U(\rv ),
\ee
where the first term represents the kinetic energy associated to the unconstrained motion in the $xy$ plane and
${\bf p}=(p_x,p_y)$ is the standard two-dimensional momentum operator.  The finite thickness $d$ of the metallic film
is taken into account by a confining potential
\be\label{VC}
V_C=V_+\theta(z-z_+)+V_-\theta(z_--z),
\ee
where $V_{\pm}$ is the height of the potential barrier at $z_{\pm}=\pm d/2$ and
$\theta(z)$ is the Heaviside function.
 The third term in Eq.(\ref{H1}) describes the
 Rashba interfacial spin-orbit interaction in the $xy$ plane located at $z_{\pm}=\pm d/2$
 \be\label{HR}
H_R=\frac{\lambda_-^2V_-\delta(z-z_-)-\lambda_+^2V_+\delta(z-z_+)}{\hbar}(p_y \sigma_x-p_x \sigma_y),
\ee
where $\lambda_{\pm}$ are the effective Compton wavelengths for the two interfaces,
$\sigma_x, \sigma_y,\sigma_z$ are the Pauli matrices.
The last term in Eq.(\ref{H1}) represents the scattering from impurities affecting the motion in the $x-y$ plane and
$\rv=(x,y)$ is the coordinate operator. The impurity potential is taken in a standard way as a white-noise disorder with
variance $\langle U(\rv )U(\rv ')\rangle = (2\pi N_0 \tau)^{-1}\delta (\rv -\rv')$,
 where $N_0$ is the two-dimensional density of states previously introduced.
We will assume throughout that the Fermi energy $E_{Fn}$ in each subband is much larger than the level broadening $\hbar /\tau$ and use the self-consistent Born approximation.

The eigenfunctions of the Hamiltonian~(\ref{H1})  have the form
 \be\label{PSI}
 \psi_{n\kv s}(\rv,z) = \frac{e ^{i\kv \cdot \rv}}{\sqrt{\cal A}} \frac{1}{\sqrt{2}}\left(
 \begin{array}{c} 1\\i s e^{i\theta_\kv} \end{array}\right) f_{n\kv s}(z),
 \ee
 where ${\cal A}$ is the area of the interface, $\kv=(k_x,k_y)$ is the in-plane wave vector,
 $\rv$ is the position in the interfacial plane and $z$ is the coordinate perpendicular to the plane.  $\theta_\kv$ is the angle between  $\kv$ and the $x$ axis.   These states are classified by a {\it subband index} $n=1,2..$, which plays the role of a principal quantum number,  an in-plane wave vector $\kv$, and an {\it helicity index},   $s=+1$ or $-1$ which determines the form of the spin-dependent part of the wave function.

By inserting the wave function (\ref{PSI}) into the Schr\"odinger equation for the Hamiltonian (\ref{H1}) we find the following equation
for the functions $ f_{n\kv s}(z)$ describing the motion along the $z$-axis
\be
\label{SC}
-\frac{\hbar^2}{2m}f''_{n\kv s}(z)+
\left\{
V_C(z) -ks\left[\lambda_-^2V_-\delta(z+d/2)-\lambda_+^2V_+\delta(z-d/2)\right]
\right\}
f_{n\kv s}(z) =  \epsilon_{n\kv s}
f_{n\kv s}(z),
\ee
where the full energy eigenvalues are
\be
\label{FEIG}
E_{n\kv s}=\frac{\hbar^2 k^2}{2m}+\epsilon_{n\kv s}.
\ee
By taking into account the continuity of the wave function $f_{n\kv s}(z)$ at $z=\pm d/2$ and the discontinuities of its derivatives
we obtain for the eigenvalue $ \epsilon_{n\kv s}$
the following transcendental equation
 \be
 \label{M1}
 \arctan{ \left( \frac{\sqrt{\epsilon}}{\sqrt{ \left(\frac{d^2}{d_-^2}-\epsilon \right)}-\frac{d}{d_-}\alpha_- sk} \right)}+\arctan{ \left( \frac{\sqrt{ \epsilon }}{\sqrt{ \left(\frac{d^2}{d_+^2}-\epsilon \right)}+\frac{d}{d_+}\alpha_+ sk}\right)}+ \sqrt{ \epsilon  } =n\pi,
 \ee
 where the energy $\epsilon$ is measured in units of 
 $E_0=\hbar^2/(2md^2)$ set by the thickness
of the film.
In the absence of spin-orbit coupling ($\lambda_{\pm}=0$) and for infinite heights of the potential ($V_{\pm}\rightarrow\infty$), the solution reduces to the well-known energy levels $\epsilon_{n\kv s}= E_0 n^2$.  
In the general case with both $\lambda_{\pm}$ and $V_{\pm}$ finite we use perturbation theory by assuming $d$ large.
There are two natural length scales associated with the confining potential $d_\pm=\hbar/\sqrt{2mV_\pm}$ so that we expand
in the small parameters $d_\pm/d$. Since all the energy scales are set by $E_0$, we find useful to describe the spin-orbit coupling
in terms of the parameters $\alpha_{\pm}=\lambda_\pm^2/d_\pm$ in such a way that the product $E_0 \alpha_{\pm}/\hbar $ has the dimensions
of a velocity, just as the typical Rashba coupling parameter. In the following we make an expansion to first order in
$d_\pm/d$ and up to  third order in  $\alpha_{\pm} k$.

For the eigenvalues of (\ref{SC}) we find

\ber
\label{E}
\epsilon_{n\kv s}=E_0n^2\left[1-2\frac{d_-+d_+}{d}+s e_1 k+e_2k^2+s e_3 k^3\right]
\eer
and the eigenfunctions
\ber
\label{f}
f_{n\kv s}(z)&=&c_{n\kv s}\sin
\left[
\frac{n\pi}{d+\frac{d_-}{1-\alpha_- ks}+\frac{d_+}{1+\alpha_+ ks}}\left(\frac{d}{2}+z+\frac{d_-}{1-\alpha_- ks}\right)
\right],
\eer
where
\ber
\nonumber
c_{n\kv s}&=&\sqrt{\frac{4}{d_e\left[2-(s e_1k +e_2k^2+s e_3k^3)\right]}}, \  d_e=d+d_++d_-;\\
e_1&=&2\left(\frac{d_+}{d}\alpha_+ -\frac{d_-}{d}\alpha_-\right), e_2=-2\left(\frac{d_+}{d}\alpha_+^2 +\frac{d_-}{d}\alpha_-^2\right), e_3=2\left(\frac{d_+}{d}\alpha_+^3 -\frac{d_-}{d}\alpha_-^3\right).\label{Cn}
\eer
Notice that the sign of the coefficients $e_1$ and $e_3$ depends on the relative strength of the spin-orbit coupling $\lambda_{\pm}$
and barrier heights $V_{\pm}$.  To avoid troubles with minus signs in the following calculations, we assume that the couplings are labeled in such a way that $\lambda_+>\lambda_-$, and $V_+>V_-$ so that $e_1, e_3>0$.

In the next Section we evaluate the SHC assuming that $n=n_c$ is the topmost occupied subband. In the following we use units such that
$\hbar=c=1$.

\section{Spin  Hall conductivity}
\label{sec_SHC}
\begin{figure}
\begin{center}
\includegraphics[width=6in]{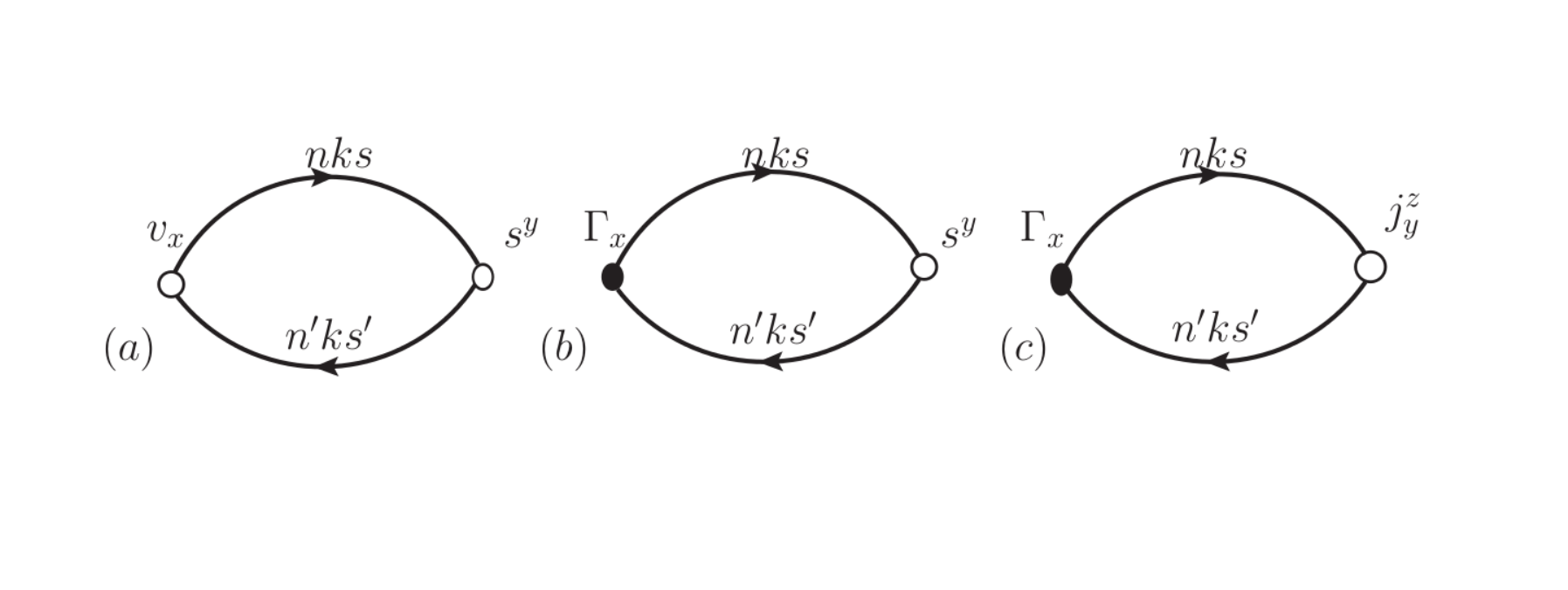}
\caption{ Feynman bubble diagram for the EC(a+b) or SHC(c). The empty right dot indicates the spin density (EC)
or the spin current density (SHC) bare vertex, the left empty one indicates the normal velocity operator, and the full dot is the dressed charge current density vertex.
}
\label{bubbles}
\end{center}
\end{figure}

The SHC is defined as the non-equilibrium spin density response to an applied electric field. By using a vector gauge with
the electric field given by ${\bf E}=-\partial_t {\bf A}$, the Kubo formula, corresponding to the bubble diagram of Fig.\ref{bubbles},
reads
\be
\label{E2_00}
\sigma^{SHE}=\lim_{\omega\rightarrow 0} \frac{{\rm Im}\langle\langle j_y^z;j_x\rangle\rangle}{\omega},
\ee
where we have introduced the spin current operator $j^z_y=\sigma_z k_y /2m$ and the charge current operator $j_x=-e{\hat v}_x$.
The number current operator, besides the standard velocity component, includes a spin-orbit induced anomalous contribution ${\hat v}_x=
k_x/m +{\hat \Gamma}_x$. Without vertex corrections, the anomalous contribution reads
\be
\label{v}
{\hat \Gamma}_x=\delta {\hat v}_x=\left[\lambda_+^2V_+\delta(z-z_+)-\lambda_-^2V_-\delta(z-z_-)\right]\sigma_y.
\ee

This expression can be written in terms of the exact Green functions and vertices as
\be
\label{E2_000}
\sigma^{SHE}=-\lim_{\omega \rightarrow 0}{\rm Im} \frac{e}{\omega}\sum_{nn'\kv \kv' s s'}\langle n'\kv' s' |\hat{v}_x|n\kv s \rangle\langle n\kv s |j_y^z|n'\kv' s' \rangle \int_{-\infty}^{\infty} \frac{d \epsilon}{2\pi} G_{ns}(\epsilon_+,\kv)G_{n's'}(\epsilon_-,\kv').
\ee
where $e>0$ is the unit charge, $\epsilon_{\pm}=\epsilon\pm \omega /2$ and $G_{n s}(\epsilon,\kv)=(\epsilon -E_{n\kv s}+{\rm i}{\rm sgn} \epsilon/2\tau)^{-1}$
is the Green function averaged over disorder in the self-consistent Born approximation with self energy
\be
\label{BA}
\Sigma_{ns} (\rv,\rv';\epsilon)=\frac{\delta (\rv -\rv')}{2\pi N_0\tau} G_{ns}(\rv,\rv;\epsilon).
\ee
After performing the integral over the frequency we obtain
\ber
\label{SH1}
\sigma^{SHE}=-\frac{e}{2\pi}\sum_{nn'\kv ss'}\langle n'\kv s' |\hat{v}_x|n\kv s \rangle\langle n\kv s |j_y^z|n'\kv s' \rangle
G^{R}_{n \kv s}G^{A}_{n' \kv s'},
\eer
where we have introduced the retarded and advanced zero-energy Green functions at the Fermi level
\be
\label{G}
G^{R,A}_{n \kv s}=\frac{1}{-E_{n\kv s}+\mu \pm {\rm i}/2\tau}
\ee
and exploited the fact that plane waves at different momentum $\kv$ are orthogonal.

To proceed further we need the expression for the vertices.
It is easy to recognize that the standard part of the velocity operator $k_x/m$ does not contribute since it requires $s=s'$, whereas
the matrix elements of $j^z_y$ differ from zero only for $s\neq s'$.  Explicitly we have
\begin{eqnarray}
\langle n'\kv s' |k_x|n\kv s \rangle
&=&
k_x\langle f_{n' \kv s'}|f_{n\kv s}\rangle\delta_{s's}=\langle f_{n' \kv s'}|f_{n\kv s}\rangle k \cos\theta_{\kv}\,\delta_{s's} \label{ME1}\\
\langle n\kv s' |\delta {\hat v}_x|n\kv s \rangle
&=&
\left(\cos\theta_{\kv}\,\sigma_{z,s's}+\sin\theta_{\kv}\,\sigma_{y,s's}\right)
\frac{\Delta E_{nk}}{k}
\langle f_{n\kv s'}|f_{n\kv s}\rangle\label{VC3}\\
\langle n\kv s |j^z_y|n'\kv s' \rangle&=& \langle f_{n\kv s}|f_{n'\kv s'}\rangle
\frac{k}{2m}\sin\theta_{\kv}\,\sigma_{x,ss'}\,,\label{ME22}
\end{eqnarray}
where $\Delta E_{nk}=(E_{nk+}-E_{nk-})/2=E_0 n^2(e_1 k+e_3k^3)$ is half the spin-splitting energy in the $n$-th band.  Eq.(\ref{VC3}) is straightforwardly obtained from the eigenvalue equation~\eqref{SC} for the functions $f_{n\kv s}(z)$.

Let us now discuss the overlaps between the wave functions $\langle f_{n\kv s}|f_{n'\kv' s'}\rangle$.
If $n=n'$ we have
\be
\label{VC10}
\langle f_{n\kv s}|f_{n\kv' s'}\rangle=\frac{d_e}{2} c_{nks} c_{nk's'}
 \left[ 1-\frac{e_1(ks+k's')+e_2(k^2+k'^2)+e_3(k^3s+k'^3s')}{4}\right],
\ee
which is unity plus corrections of order $(d_{\pm}/d)$ when $s ,k \neq s',k'$.
If $n \neq n'$ $\langle f_{n\kv s}|f_{n'\kv' s'}\rangle$ is at least of order $(d_{\pm}/d)$.
Before  continuing our calculation we observe that it is important to distinguish between the intra-band
($n=n'$) and the inter-band ($n\neq n'$) contributions.
The inter-band contributions are of second order in $d_\pm/d$,
because they are proportional to $\langle f_{n\kv s}|f_{n'\kv s'}\rangle ^2$.  Since we  limit our expansion to the first order in $d_\pm/d$ we will from now on neglect these contributions. 
Notice, however, that this approximation is no longer valid when the intra-band splitting controlled by $e_1$ and $e_3$ vanishes.
In this case one cannot avoid taking into account the inter-band contributions.
In the same spirit, we also approximate
the intra-band overlap $\langle f_{n\kv s}|f_{n\kv' s'}\rangle \simeq 1$, because all of our results are at least linear in $(d_{\pm}/d)$ and we neglect higher order terms.

\begin{figure}
\begin{center}
\includegraphics[width=4.5in]{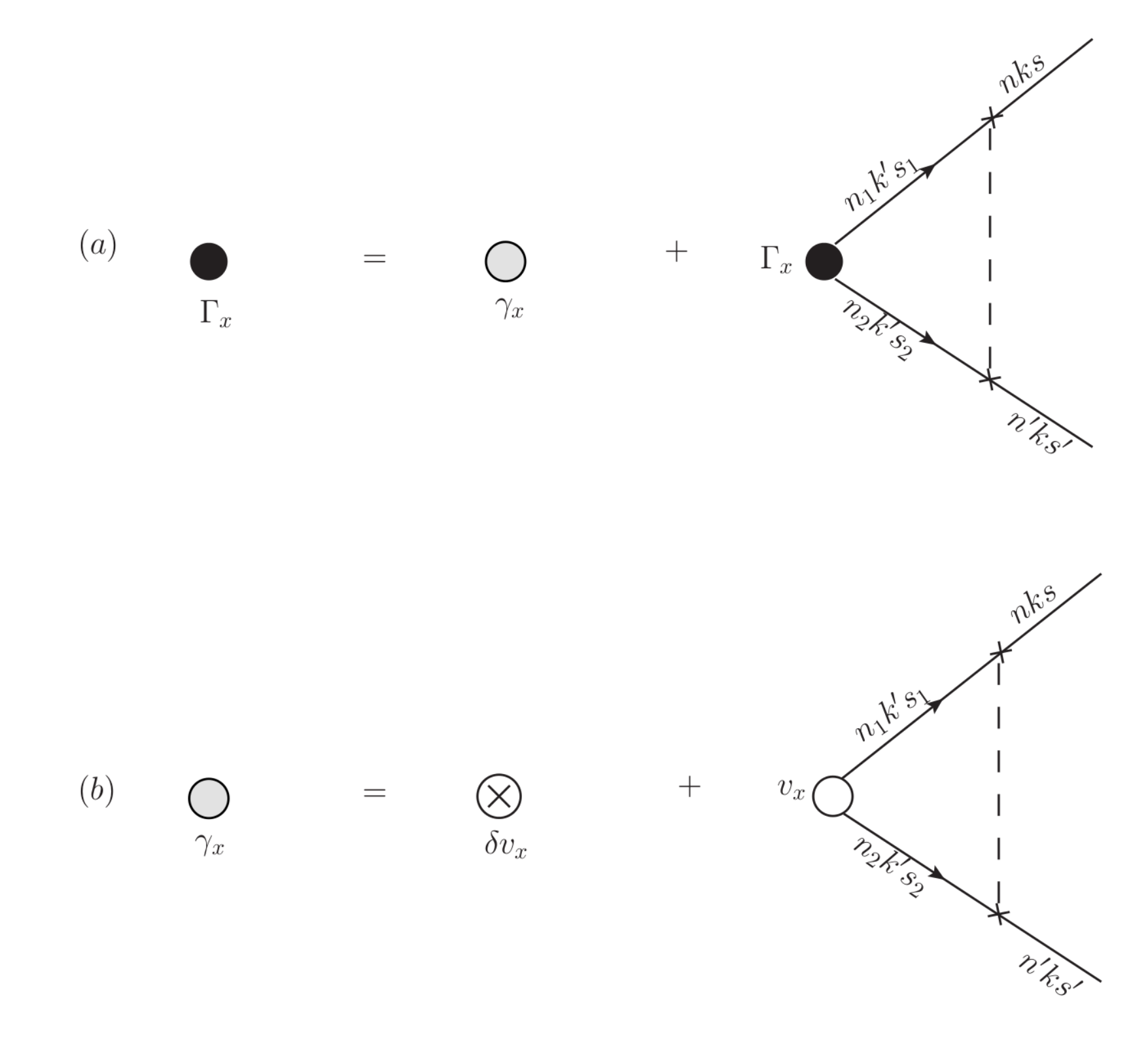}
\caption{Ladder resummation for the spin-dependent part of the dressed charge current density vertex. The dashed line represents the correlation between propagators scattering off the same impurity site.}
\label{vertex}
\end{center}
\end{figure}

The anomalous contribution to the velocity vertex,  ${\hat \Gamma}_x$, can be computed
following the procedure described in Ref.~\onlinecite{Raimondi05}
 according to the equations (see Fig.\ref{vertex})
 \ber
\label{VC1}
{\hat \Gamma}_x &=&{\tilde \gamma}_x +\frac{1}{2\pi N_0 \tau}\sum_{\kv'} G^R_{\kv'}{\hat \Gamma}_xG^A_{\kv'}\,, \nn\\
{\tilde \gamma}_x&=& {\hat \delta v}_x+\frac{1}{2\pi N_0 \tau}\sum_{\kv'} G^R_{\kv'}\ \frac{k'_x}{m} \ G^A_{\kv'}\equiv {\tilde \gamma}^{(1)}+{\tilde \gamma}^{(2)}
\eer

%
To extend the treatment to the present case, the projection must be made over the states $|n\kv s \rangle$.
Assuming that the impurity potential does not depend on $z$, the matrix elements of the effective vertex ${\tilde \gamma}^{(2)}$ are:
\be
\label{VC2}
\gamma^{(2)nn}_{ss'}(k)\equiv
\langle n\kv s |{\tilde \gamma}^{(2)}|n\kv s' \rangle=
\frac{1}{2\pi N_0 \tau}\sum_{n_1 \kv's_1} \langle n\kv s  |n_1\kv' s_1 \rangle G^R_{n_1 \kv's_1} \ \frac{k'_x}{m} \ G^A_{n_1\kv's_1}\langle n_1 \kv' s_1   |n \kv s' \rangle,
\ee
and $\gamma^{(1)nn}_{ss'}(k)\equiv
\langle n\kv s |{\tilde \gamma}^{(1)}|n\kv s' \rangle$ is given by Eq.(\ref{VC3}).
The matrix elements $\langle n \kv s  |n_1 \kv' s_1 \rangle$ and $\langle n_1 \kv' s_1   |n \kv s' \rangle$ are those of the impurity potential:
\begin{eqnarray}
\langle n \kv s  |n_1 \kv' s_1 \rangle&=& \frac{1}{2}
\langle f_{n\kv s}|f_{n_1\kv' s_1}\rangle \left[ 1+s s_1 e^{{\rm i}(\theta_{\kv'}-\theta_{\kv})}\right]\\
\langle n_1 \kv' \lambda_1   |n \kv s' \rangle&=&
\frac{1}{2}
\langle f_{n_1\kv's_1}|f_{n\kv s'}\rangle \left[ 1+s' s_1 e^{-{\rm i}(\theta_{\kv'}-\theta_{\kv})}\right].
\end{eqnarray}
By observing that $k_x'=k' \cos \theta_{\kv'}$, one can perform the integration over the direction of $\kv'$ in  the expression of $\gamma^{(2)nn}_{ss'}(k)$
\begin{equation}
\label{VC7}\frac{1}{4}
\int_0^{2\pi} \frac{{\rm d}\theta_{\kv'}}{2\pi}
\left[ 1+s s_1 e^{{\rm i}(\theta_{\kv'}-\theta_{\kv})}\right]
\cos\theta_{\kv'}\,
\left[ 1+s' s_1 e^{-{\rm i}(\theta_{\kv'}-\theta_{\kv})}\right]=\frac{s_1}{8}\left[ s e^{-{\rm i}\theta_{\kv}}+
s'  e^{{\rm i}\theta_{\kv}}\right],
\end{equation}
to get
\be
\label{VC8}
\gamma^{(2)nn}_{ss'}(k)=
\frac{(\cos\theta_{\kv}\,\sigma_{z,ss'}+\sin\theta_{\kv}\,\sigma_{y,ss'})}{16\pi N_0 \tau}\sum_{n_1 \kv' s_1}s_1
\langle f_{n\kv s}|f_{n_1\kv' s_1}\rangle
\langle f_{n_1\kv' s_1}|f_{n\kv s'}\rangle
 G^R_{n_1 \kv' s_1} \ \frac{ k'}{m} \ G^A_{n_1\kv' s_1}.
\ee

Approximating  $\langle f_{n\kv s}|f_{n_1\kv' s_1}\rangle \sim \delta_{nn_1}$, summing over $s_1$, and integrating over $k$ with the technique shown in the Appendix  yields
\be
\label{VC13}
\gamma^{(2)nn}_{ss'}(k)=-(\cos\theta_{\kv}\,\sigma_{z,ss'}+\sin\theta_{\kv}\,\sigma_{y,ss'})
E_0 n^2 (e_1+2e_3 k_{Fn}^2)\,,
\ee
where we  have introduced the spin-averaged Fermi momentum in the $n$-th subband
\ber
\label{VC16}
\frac{k_{Fn}^2}{2m}&=& \mu-E_0n^2.
\eer
{ On the other hand $\gamma^{(1)nn}_{ss'}(k)$ is given by
\be
\gamma^{(1)nn}_{ss'}(k)=(\cos\theta_{\kv}\,\sigma_{z,ss'}+\sin\theta_{\kv}\,\sigma_{y,ss'})
E_0 n^2 (e_1+e_3 k_{Fn}^2)
\ee
where $k$ has been replaced by $k_{Fn}$ at the required level of accuracy.
Combining $\gamma^{(1)nn}_{ss'}(k)$ and $\gamma^{(2)nn}_{ss'}(k)$ as mandated by Eq.~(\ref{VC1})  we finally obtain}
\be
\gamma^{nn}_{x,ss'}(k)=
-(\cos\theta_{\kv}\, \sigma_{z,ss'}+\sin\theta_{\kv}\, \sigma_{y,ss'})E_0 n^2 e_3k_{Fn}^2\,.
 \label{VC14}
\ee
Next  we  project the equation for the  vertex corrections in the  basis of the eigenstates and get the following integral equation:
\be
\label{VC18}
\Gamma^{nn}_{x,ss'}(k)=\gamma^{nn}_{x,ss'}(k)+
\frac{1}{2\pi N_0\tau}\sum_{n_1n_2\kv' s_1s_2}\langle n\kv s|n_1\kv' s_1 \rangle G^R_{n_1\kv' s_1} \Gamma^{n_1n_2}_{x,s_1s_2}(k') G^A_{n_2\kv' s_2}\langle n_2\kv' s_2|n\kv s' \rangle,
\ee
which, by confining to intra-band processes only,  can be solved with  the ansatz $\Gamma^{nn}_{x,ss'}(k) =\Gamma^{n}(k_{Fn}) (\cos (\theta_{\kv}) (\sigma_z)_{ss'}+\sin (\theta_{\kv}) (\sigma_y)_{ss'})$ yielding
\be
\label{VC19}
\Gamma^{nn}_{x,ss'}(k)=\gamma^{nn}_{x,ss'}(k)\frac{\tau_{DP}^{(n)}}{\tau}.
\ee


By performing the integral over momentum and summing over the spin indices in Eq.(\ref{SH1}), one obtains  the SHC as
\be
\label{SH3}
\sigma^{SHE}=\sum_{n=1}^{n_c}\frac{e}{8 \pi}   \frac{2\tau}{\tau_{DP}^{(n)}}\frac{\Gamma^n (k_{Fn})}{\Delta E_{nk_{Fn}}/k_{Fn}},
\ee
where $n_c$ is the number of occupied bands.

If  vertex corrections are ignored, i.e.,  if we approximate  $\Gamma^n (k_{Fn})=\Delta E_{nk_{Fn}}/k_{Fn}$ (cf. Eq.(\ref{VC3})),
Eq.(\ref{SH3}) gives us
\be
\label{SH3drift}
\sigma^{SHE}_{drift}=\sum_{n=1}^{n_c}\frac{e}{8 \pi}   \frac{2\tau}{\tau_{DP}^{(n)}}\,,
\ee
which,  in the weak disorder limit ($\tau \rightarrow\infty$),   reproduces the
result of Ref.  \onlinecite{Wang2013}, {\it i.e.}  $\sigma^{SHE}_{drift}=(e/8\pi) n_c$.

If instead the renormalized vertex (\ref{VC19}) is properly taken into account, we obtain
\begin{equation}
\label{SH5}
\sigma^{SHE}=-\sum_{n}^{n_c}\frac{e}{4\pi} \frac{e_3k_{Fn}^2}{ e_1+e_3k_{Fn}^2}.
\end{equation}
Notice that, being proportional to $\lambda_{\pm}^4$ ($e_1\propto \lambda_{\pm}^2$, $e_3\propto \lambda^6_{\pm}$), this result is consistent with the result obtained in Ref.~\onlinecite{Hayden13} for a different but related model.
Making use of the explicit expressions for  $e_1$ and $e_3$ we finally get the previously reported result of Eq.(\ref{SH4-Preview}).

\section{Edelstein conductivity}
\label{sec_EC}

In the d.c. limit, i.e., for $\omega\rightarrow 0$, the Edelstein conductivity (EC) is defined by

\be
\label{E2_0}
\sigma^{EE}=\lim_{\omega\rightarrow 0} \frac{{\rm Im}\langle\langle s^y;j_x\rangle\rangle}{\omega}.
\ee

That can be written as:
\be
\label{E2}
\sigma^{EE}=-\lim_{\omega \rightarrow 0}{\rm Im} \frac{e}{\omega}   \sum_{n n' \kv\kv'  s s'}\langle n'\kv' s' |{\hat v}_x|n\kv s \rangle\langle n\kv s |s^y|n'\kv' s' \rangle \int^{\infty}_{-\infty} \  \frac{d \epsilon}{2\pi} G_{n s}(\epsilon_+,\kv)G_{n' s'}(\epsilon_-,\kv'),
\ee

After performing the integral over  frequency we get
\ber
\label{E3}
\sigma^{EE}=-\frac{e}{2\pi}\sum_{n n' \kv s s'}\langle n'\kv s' |{\hat v}_x|n\kv s \rangle\langle n\kv s |s^y|n'\kv s' \rangle
G^{R}_{n \kv s}G^{A}_{n' \kv s'}\,,
\eer
where we have used again the orthogonality of the eigenvectors with different momentum.
As shown in Fig.\ref{bubbles}, we consider the bare vertex for the spin
density  $s^y=\sigma_y/2$ and the two vertices for the number current density
${\hat v}_x=\hat{\Gamma}_x + k_x/m$,\cite{Raimondi05} -- $\hat{\Gamma}_x$ being the renormalized spin-dependent part of the  vertex.
Clearly, the two parts of the number current vertex yield two separate contributions to the EC and we are now going to evaluate them
separately.
We then evaluate the (a) diagram in Fig.\ref{bubbles} as:
\ber
\label{E4}
\sigma^{EE,(a)}=-\frac{e}{4\pi m}\sum_{nn' \kv ss'}\langle n'\kv s' |k_x|n\kv s \rangle\langle n\kv s |\sigma_y|n'\kv s' \rangle
G^{R}_{n \kv s}G^{A}_{n' \kv s'},
\eer
where the matrix elements of the spin vertex is
\begin{eqnarray}
\langle n\kv s |\sigma_y|n'\kv s' \rangle
&=& \langle f_{n\kv s}|f_{n'\kv s'}\rangle
(\cos\theta_{\kv}\,\sigma_{z,ss'}-\sin\theta_{\kv}\,\sigma_{y,ss'}).\label{ME2}
\end{eqnarray}

Setting  $n'=n$ and  using Eq.(\ref{E}) for the energy eigenvalues,  we can
perform the integration over the momentum in Eq.(\ref{E4}) obtaining for
 $\sigma^{EE,(a)}$ the expression
\ber
\label{E5}
\sigma^{EE,(a)}=\sum_{n=1}^{n_c}eN_0\tau E_0n^2\left(e_1+2e_3k_{Fn}^2\right),
\eer


Next we evaluate  the (b) diagram in Fig.\ref{bubbles} as:
\ber
\label{E10}
\sigma^{EE,(b)}=-\frac{e}{4\pi}\sum_{nn' \kv s s'}  \langle n'\kv s' |{\hat \Gamma}_x|n\kv s \rangle\langle n\kv s |\sigma_y|n'\kv s' \rangle
G^R_{n\kv s}G^A_{n'\kv s'},
\eer


{  We set $n=n'$ and insert the result obtained in Eq.(\ref{VC19}) for $\langle n\kv s' |{\hat \Gamma}_x|n\kv s \rangle$}. 
Since both the matrix elements of ${\hat \Gamma}_x$ and $\sigma_y$ contain  terms proportional to $\cos (\theta_{\kv})$ and
$\sin (\theta_{\kv})$,
 we  must distinguish between $s=s'$ (first term in Eq.(\ref{VC14})) and $s\neq s'$
(second term in Eq.(\ref{VC14})).  
If  $s=s'$  we have
\ber
\label{E10b}
\sigma^{EE,(b)}_{1}=-\frac{e}{4\pi}\sum_{n \kv s } \langle n s |{\tilde \Gamma}_x|n\kv s \rangle\langle n\kv s |\sigma_y|n \kv s \rangle
G^R_{n\kv s}G^A_{n\kv s}
\eer
The integral over the momentum can be done with the technique shown in the Appendix to yield
\ber
\label{E6}
\sigma^{EE,(b)}_{1}=\sum_{n}^{n_c}eN_0\tau E_0n^2e_3k_{Fn}^2
\frac{\tau_{DP}^{(n)}}{2\tau}.
\eer
If $s\neq s'$ we have instead
\ber
\label{E7}
\sigma^{EE,(b)}_{2}=-\frac{e}{4\pi}\sum_{n\kv  s}\langle n \kv \bar s |{\tilde \Gamma}_x|n\kv s \rangle\langle n\kv s |\sigma_y|n \kv' \bar s \rangle
G^R_{n\kv s}G^A_{n\kv \bar s}.
\eer
So we can conclude that
\ber
\label{E8}
\sigma^{EE,(b)}_{2}=\sum_{n=1}^{n_c}eN_0\tau E_0n^2\frac{e_3k_{Fn}^2}{( 2 \tau \Delta E_{nk_{Fn}}  )^2}
\eer
with $\Delta E_{nk_{Fn}}$ defined in Eq.(\ref{deltaE}). 
Combining the (a) and (b) contributions, the final result for the Edelstein conductivity  is found to be:
\ber
\label{E9}
\sigma^{EE}=\sum_{n=1}^{n_c}eN_0 \tau E_0 n^2 \left[e_1+3e_3 k_{Fn}^2+
\frac{2 e_3 k_{Fn}^2}{ ( 2 \tau \Delta E_{nk_{Fn}}  )^2}\right]\,,
\eer
which is easily seen to be equivalent to Eq.~(\ref{E9-Preview}).


\section{Discussion}
\label{sec_discussion}

The two central results \eqref{E9} and \eqref{SH5} may be interpreted along the lines outlined in the introduction.
We begin by noticing that both conductivities are expressed as simple sums of independent subband contributions,  
hence the relation \eqref{discussion_3} is valid separately within each subband.  
The second step is the identification of the quantity $\tau_s /L_{so}$ for a given subband.
Clearly $\tau_s$ must be identified with the Dyakonov-Perel relaxation time $\tau_{DP}^{(n)}$ 
defined in \eqref{a}.  For the spin-orbit length $L_{so}$ one notices
that the quantity $2\alpha p_F$ in the Rashba model corresponds to the band splitting,
and hence must here be replaced by $-2\Delta E_{nk_{Fn}}$.  This yields,
after restoring $\hbar$ in the following,
\begin{equation}
\label{interpretation_2}
L_{so}^{(n)}=\frac{\hbar v_{Fn}}{2\Delta E_{nk_{Fn}}},
\end{equation}
i.e. $\tau_s/L_{so} \rightarrow \tau_{DP}^{(n)}/L_{so}^{(n)}$.
With this prescription one can apply Eq.~\eqref{discussion_3} subband-by-subband and obtain
\be
\sigma^{EE,(n)}=\frac{\tau_{DP}^{(n)}}{L_{so}^{(n)}}
\left[
\sigma^{SHE,(n)}-\sigma^{SHE,(n)}_{drift}
\right],
\ee
where $\sigma^{SHE,(n)}, \sigma^{SHE,(n)}_{drift}$ stand for the $n$-th band contribution to
Eqs.~\eqref{SH5} and \eqref{SH3drift}, respectively.  It is now immediate to see
that a sum over the subbands leads to the EC of Eq.~\eqref{E9}.
We may thus conclude the following: a non vanishing SHC in the presence of Rashba spin-orbit coupling gives rises
to an anomalous EC scaling with the inverse scattering time; conversely, 
an anomalous EC yields a non-vanishing SHC.


We now consider two physically interesting limiting cases of the general solution:
\begin{enumerate}
\item the insulator-metal-vacuum junction, $\lambda_+=0$ $V_+\rightarrow\infty$, $\lambda_-=\lambda$ $V_-=V$;
\item films with the same spin orbit constant coupling at the two interfaces, $\lambda_-=\lambda_+=\lambda$.
\end{enumerate}
In the first case we get
\ber
\label{V1}
\sigma^{EE}=-\sum_{n}^{n_c}\frac{2eN_0 \tau E_0 n^2 \lambda^2}{d\hbar } \left(1+\frac{\hbar^2\pi^2 V}{8 \tau^2  E_0^3  n^4}\right),
\eer
\begin{equation}
\label{V2}
\sigma^{SHE}=-\sum_{n}^{n_c}\frac{e}{4\pi\hbar^3 } 2mk_{Fn}^2V\lambda^4.
\end{equation}
There are some experimental studies of metal-metal-vacuum junctions that shows giant spin-orbit 
coupling\cite{Rybkin2010} and where one could test the prediction of Eqs.(\ref{V1}-\ref{V2}).
Though Eq.~\eqref{V2} is obtained for small values of the parameter $2mk_{Fn}^2V\lambda^4/\hbar^2 \ll1$,
the structure of the result is quite interesting: it suggests that this kind of device, 
the insulator-metal-vacuum junction, could be an efficient spintronic device, its transport properties 
being proportional to the barrier height $V$.

In the second case let us first assume a ``quasi-symmetric'' configuration,
i.e. though $\lambda_+=\lambda_-\equiv\lambda$, the barrier heights are different, $V_+\neq V_-$.
We then obtain that the spin splitting of the bands vanishes to linear order in $k$ 
($e_1=0$){ (see footnote 48)} so that
\begin{equation}  \label{V4}
\sigma^{SHE}=-\sum_{n}^{n_c}\frac{e}{4\pi\hbar }\,,
\end{equation}
and
\ber
\label{V3}
\sigma^{EE}=\sum_{n}^{n_c}eN_0 \tau \frac{\Delta E_{nk_{Fn}}}{k_{Fn}\hbar }
\left[3+\frac{\hbar^2}{2  (\tau\Delta E_{nk_{Fn}})^2}\right]\,.\eer
The SHC  in this limit is independent of $\lambda$.  
This very striking result is reminiscent of the universal result $\frac{e}{8\pi\hbar}$ 
obtained for a single Bychkov-Rashba band when vertex corrections are ignored.~\cite{Sinova04}
However vertex corrections are now fully included, yet the SHC is not only finite, 
but {\it independent} of $\lambda$ and equal to the single band universal result multiplied by a factor $-2$!  
We emphasize that this result has nothing to do with the non-vanishing intrinsic SHC that arises 
in certain generalized models of spin-orbit coupling with winding number higher than 1.~\cite{Engel05} 
Rather, it has everything to do with the $k$-dependence of the transverse subbands describing 
the electron wave function in the $z$- direction. 
We also find that the anomalous part of the Edelstein effect becomes large, 
as it is proportional to $1/\Delta E_{nk_{Fn}}$, and the splitting vanishes 
with the third power of $k$ at small $k$.

Let us finally discuss the fully inversion-symmetric limit of the model, $\lambda_+=\lambda_-$ and $V_+=V_-$. 
We notice that in this case the limit of Eq.~\eqref{SH5} does not exist, 
because both $e_1$ and $e_3$ vanish (the spin splitting is identically zero!) 
while the value of Eq.~\eqref{SH5} depends on the order in which $e_1$ and $e_3$ tend to zero, 
in particular on whether they tend to zero simultaneously, or $e_1$ tend to zero before $e_3$, 
as in the ``quasi-symmetric'' case above.
The origin of this apparently unphysical non-analytic behavior can be traced back to the singular character 
of the vertex \eqref{VC19} for vanishing spin splitting.
Under these circumstances, the Dyakonov-Perel spin relaxation time \eqref{a} diverges, apparently implying
spin conservation.  However, even in the inversion-symmetric limit, interband effects provide spin relaxation 
processes which regularize the vertex.  Such effects are typically negligible away from the inversion-symmetric 
limit, since they are proportional to the square of the wave-function overlap 
between different bands and therefore scale as $(d_{\pm}/d)^2$.  However, in the inversion-symmetric limit they cannot be neglected.

A full analysis of interband effects is beyond the scope of the present paper, and we limit ourselves 
to a heuristic discussion of the physical origin of the spin relaxation mechanism due to 
interband virtual transitions.
In the inversion-symmetric limit, the Hamiltonian is invariant upon the simultaneous operations of space inversion along the
$z$-direction ($z\rightarrow -z$) and helicity flipping ($s\rightarrow -s$), i.e., a full mirror reflection in the $x-y$ plane.  Hence the eigenfunctions can be classified as even or odd under such a reflection:
\be
\label{symmetry}
f_{n\kv s}(z)=P_n f_{n\kv-s} (-z)\,.
\ee
where $P_n=\pm 1$.   Furthermore the parity eigenvalue $P_n$ is the same as in the absence of spin-orbit interaction, because  the reflection commutes with the spin-orbit interaction.

Since states of opposite helicity are degenerate, one can construct,
in each band $n$, states that are  linear combinations of the helicity eigenstates $|\pm\rangle$
\ber
\psi_{n\kv \uparrow}&=&\frac{1}{2}\left( f_{n\kv +}(z)|+\rangle + f_{n\kv -}(z)|-\rangle\right) \\
\psi_{n\kv \downarrow}&=&\frac{1}{2}\left( f_{n\kv +}(z)|+\rangle - f_{n\kv -}(z)|-\rangle\right)\,.
\eer
These can be rewritten in terms of the eigenstates $|\uparrow \rangle$ and $|\downarrow \rangle$ of $\sigma_z$
and, after using (\ref{symmetry}), one obtains
\ber
\psi_{n\kv \uparrow}&=&\frac{f_{n\kv +}(z)+P_nf_{n\kv +}(-z)}{2}|\uparrow\rangle +{\rm i}e^{{\rm i}\theta_{\kv}} \frac{f_{n\kv +}(z)-P_nf_{n\kv +}(-z)}{2}|\downarrow\rangle \\
\psi_{n\kv \downarrow}&=&\frac{f_{n\kv +}(z)-P_nf_{n\kv +}(-z)}{2}|\uparrow\rangle +{\rm i}e^{{\rm i}\theta_{\kv}}  \frac{f_{n\kv +}(z)+P_nf_{n\kv +}(-z)}{2}|\downarrow\rangle.
\eer

One sees immediately that, within the first Born approximation, impurity
scattering cannot produce spin flipping within a band because the
matrix element of the $z$-independent disorder potential between
$\psi_{n\kv\up}$ $\psi_{n\kv'\down}$   vanishes by symmetry.

On the other hand, spin flipping may occur in the second Born approximation
by going through an intermediate state in a band of opposite parity. For
example,   an electron may first jump, under the action of the disorder
potential, to a state of {\it opposite} spin in an unoccupied band of
opposite parity; then in a second step  it may return to the original band
without flipping its spin. Alternatively the spin may remain unchanged in
the transition to the unoccupied band, and flip on the way back to the
original band.
As a result of such second-order processes, a new mechanism of spin
relaxation arises, which we call {\it inter-band spin relaxation},
with rate $\tau_{IB}^{-1}$.  When this additional relaxation mechanism is
taken into account, the diverging DP relaxation time in Eq.~(\ref{VC19}) for
the vertex is replaced by the finite total spin relaxation time
$(\tau_{DP}^{-1}+\tau_{IB}^{-1})^{-1}$.  Thus, the non-analyticity is cured.

The regime analyzed in this paper corresponds to the situation in which
$\tau_{DP}^{-1}\gg \tau_{IB}^{-1}$, and inter-band spin relaxation can be
neglected. Clearly, when looking
at the fully symmetric limit, with vanishing spin splitting, inter-band
relaxation must be taken into account,
together with inter-band contributions to the SHC and EC.  Once more, a
full-fledged treatment
of this regime is beyond the scope of the present work.


\section{Conclusions}
\label{sec_conclusions}

We have developed a simple model for describing spin transport effects and spin-charge conversion 
in heterostructures consisting of a metallic film sandwiched between two different insulators.  
All the effects we have considered depend crucially on the three-dimensional nature of the system 
-- in particular, the fact that the transverse wave functions depend on the in-plane momentum -- 
and on the lack of inversion symmetry caused by the different properties of the top and bottom metal-insulator
interfaces, each characterized by a different barrier height (gap) and spin-orbit coupling strength.
After a careful consideration of vertex corrections we find that the model supports a non-zero intrinsic SHC, 
in sharp contrast to the 2DEG Rashba case.  Strikingly, in a ``quasi-symmetric'' junction the SHC
reaches a maximal and universal value.
We have also calculated the Edelstein effect for the same model and found that the induced spin polarization
is the sum of two different contributions.  The first one is analogous to the term found in the 2DEG Rashba case,  
whereas the second ``anomalous'' one has a completely different nature.  Namely, it is inversely proportional to the scattering time, 
indicating that it is caused by the combined action of multiple electron-impurity scattering and spin-orbit coupling.  
We have also discussed the general connection between the non-vanishing SHC and the anomalous term in the EC.  
Furthermore, by Onsager's reciprocity relations, our results are immediately relevant 
to the inverse Edelstein effect\cite{Shen2013,Ganichev2002,RojasSanchez2013}, 
in which a non-equilibrium spin density induces a charge current.  
The above features, although discussed here for a specific model, are expected to be general,
proper to any non-strictly two-dimensional system in which the spin-orbit interaction is non-homogeneous 
across the confining direction.  
Technical applications of this idea could lead to a new class of spin-orbit-coupling-based devices.

\begin{acknowledgements}
CG acknowledges support  by CEA through the DSM-Energy Program (project E112-7-Meso-Therm-DSM).
GV acknowledges support from NSF Grant No. DMR-1104788.
\end{acknowledgements}



\appendix
\section{Integrals of Green functions}
To perform the integral of Eq.(\ref{E4}) we  exploit the poles with the Cauchy theorem of residues.
We use the formulae \cite{Raimondi05}
\ber
\sum_{\kv} G^R_{n\kv s}G^A_{n\kv s} f(\kv)&=&2\pi N_{ns} \tau f(k_{Fns}),\\
\sum_{\kv} G^R_{n\kv -}G^A_{n\kv +} f(\kv)&=&\frac{2\pi N_{0} \tau}{1-{\rm i}2\tau \Delta E_{nk_{F_n}}} f(k_{F_n}),
\eer
where $f(\kv)$ is assumed to be regular, $N_{ns}$ is the density of states in the $n$-subband and $k_{Fns}$ is the
corresponding momentum.
Following Ref. \onlinecite{Hayden13} the expression for both the density of states
and the Fermi momentum can be obtained in terms of the coefficients of the energy eigenvalues expansion
\ber
k_{Fns}&=&  k_{Fn}+s\frac{e_1}{2}-\frac{e_1^2}{8k_{Fn}}-s\left(\frac{e_1e_2}{2}-\frac{e_3 k_{Fn}^2}{2}\right) \\
N_{ns}&=& N_0 \left( 1+s \frac{e_1}{2k_{Fn}}-e_2+s\left(\frac{e_1e_2}{k_{Fn}}-\frac{3e_3k_{Fn}}{2}-\frac{e_1^3}{16k_{Fn}^3} \right)\right).
\eer
Hence, for instance,
\be
\sum_{\kv s} s G^R_{n\kv s} G^A_{n\kv s} k =2\pi \tau \sum_s   s k_{Fns}N_{ns}=E_0 n^2(e_1+2e_3k_{Fn}^2).
\ee

\input{v9.bbl}
\end{document}

%% file: v9.bbl
%